\documentclass[12pt]{article}
\usepackage{epsfig,graphicx}
\usepackage{fpcp04}

\def\beq{\begin{equation}}
\def\eeq#1{\label{#1}\end{equation}}
\def\eeqn{\end{equation}}

\def\beqa{\begin{eqnarray}}
\def\eeqa#1{\label{#1}\end{eqnarray}}
\def\eeqan{\end{eqnarray}}

\let\bar=\overbar

\def\Dslash{\not{\hbox{\kern-4pt $D$}}}
\def\dslash{\not{\hbox{\kern-2pt $\del$}}}

\def\msb{{\bar{\ssstyle M \kern -1pt S}}}

\def\BB0bar{B^0 {\overline B}^0}
\def\BB0dbar{B_d^0 {\overline B}_d^0}
\def\BB0sbar{B_s^0 {\overline B}_s^0}

\RequirePackage{xspace}

\usepackage{relsize}
\def\babar{\mbox{\slshape B\kern-0.1em{\smaller A}\kern-0.1em
    B\kern-0.1em{\smaller A\kern-0.2em R}}}

\def\ee         {\ensuremath{e^-e^-}\xspace}

\def\g     {\ensuremath{\gamma}\xspace}

\def\Kbar  {\kern 0.2em\overline{\kern -0.2em K}{}\xspace}

\def\Kz    {\ensuremath{K^0}\xspace}
\def\Kzb   {\ensuremath{\Kbar^0}\xspace}
\def\KzKzb {\ensuremath{\Kz \kern -0.16em \Kzb}\xspace}
\def\Kp    {\ensuremath{K^+}\xspace}
\def\Km    {\ensuremath{K^-}\xspace}

\def\KpKm  {\ensuremath{\Kp \kern -0.16em \Km}\xspace}

\def\Dbar    {\kern 0.2em\overline{\kern -0.2em D}{}\xspace}

\def\Dz      {\ensuremath{D^0}\xspace}
\def\Dzb     {\ensuremath{\Dbar^0}\xspace}
\def\DzDzb   {\ensuremath{\Dz {\kern -0.16em \Dzb}}\xspace}
\def\Dp      {\ensuremath{D^+}\xspace}
\def\Dm      {\ensuremath{D^-}\xspace}

\def\DpDm    {\ensuremath{\Dp {\kern -0.16em \Dm}}\xspace}

\def\Bbar    {\kern 0.18em\overline{\kern -0.18em B}{}\xspace}

\def\BB      {\ensuremath{B\Bbar}\xspace} 
\def\Bz      {\ensuremath{B^0}\xspace}
\def\Bzb     {\ensuremath{\Bbar^0}\xspace}
\def\BzBzb   {\ensuremath{\Bz {\kern -0.16em \Bzb}}\xspace}
\def\Bu      {\ensuremath{B^+}\xspace}
\def\Bub     {\ensuremath{B^-}\xspace}

\def\BpBm    {\ensuremath{\Bu {\kern -0.16em \Bub}}\xspace}

\mathchardef\Upsilon="7107
\def\Y#1S{\ensuremath{\Upsilon{(#1S)}}\xspace}

\mathchardef\Deltares="7101
\mathchardef\Xi="7104
\mathchardef\Lambda="7103
\mathchardef\Sigma="7106
\mathchardef\Omega="710A

\def\Deltabar{\kern 0.25em\overline{\kern -0.25em \Deltares}{}\xspace}
\def\Lbar{\kern 0.2em\overline{\kern -0.2em\Lambda\kern 0.05em}\kern-0.05em{}\xspace}
\def\Sigbar{\kern 0.2em\overline{\kern -0.2em \Sigma}{}\xspace}
\def\Xibar{\kern 0.2em\overline{\kern -0.2em \Xi}{}\xspace}
\def\Obar{\kern 0.2em\overline{\kern -0.2em \Omega}{}\xspace}
\def\Nbar{\kern 0.2em\overline{\kern -0.2em N}{}\xspace}
\def\Xb{\kern 0.2em\overline{\kern -0.2em X}{}\xspace}

\newcommand{\tev}{\ensuremath{\mathrm{\,Te\kern -0.1em V}}\xspace}
\newcommand{\gev}{\ensuremath{\mathrm{\,Ge\kern -0.1em V}}\xspace}
\newcommand{\mev}{\ensuremath{\mathrm{\,Me\kern -0.1em V}}\xspace}
\newcommand{\kev}{\ensuremath{\mathrm{\,ke\kern -0.1em V}}\xspace}
\newcommand{\ev}{\ensuremath{\mathrm{\,e\kern -0.1em V}}\xspace}
\newcommand{\gevc}{\ensuremath{{\mathrm{\,Ge\kern -0.1em V\!/}c}}\xspace}
\newcommand{\mevc}{\ensuremath{{\mathrm{\,Me\kern -0.1em V\!/}c}}\xspace}
\newcommand{\gevcc}{\ensuremath{{\mathrm{\,Ge\kern -0.1em V\!/}c^2}}\xspace}
\newcommand{\mevcc}{\ensuremath{{\mathrm{\,Me\kern -0.1em V\!/}c^2}}\xspace}

\def\mus  {\ensuremath{\rm \,\mus}\xspace}

\def\mus        {\ensuremath{\,\mu{\rm s}}\xspace}    

\def\to                 {\ensuremath{\rightarrow}\xspace}

\def\pep2{PEP-II}

\def\gsim{{~\raise.15em\hbox{$>$}\kern-.85em
          \lower.35em\hbox{$\sim$}~}\xspace}
\def\lsim{{~\raise.15em\hbox{$<$}\kern-.85em
          \lower.35em\hbox{$\sim$}~}\xspace}

\xspace

\def\jetset74   {\mbox{\tt Jetset \hspace{-0.5em}7.\hspace{-0.2em}4}\xspace}

\def\be{\begin{equation}}
\def\ee{\end{equation}}
\def\ba{\begin{eqnarray}}
\def\ea{\end{eqnarray}}
\def\bq{\begin{quotation}\noindent}
\def\eq{\end{quotation}}

\def\to{\rightarrow}
\def\nn{\nonumber\\}

\def\PR#1#2#3 {{\it Phys. Rev. }{\bf D#1} #2 {(#3)}}
\def\PRL#1#2#3 {{\it Phys. Rev. Lett. }{\bf #1} #2 {(#3)}}
\def\PL#1#2#3 {{\it Phys. Lett. }{\bf #1} #2 {(#3)}}
\def\AP#1#2#3 {{\it Ann, Phys. }{\bf #1} #2 {(#3)}}
\def\ZP#1#2#3 {{\it Z. Phys. }{\bf #1} #2 {(#3)}}
\def\NP#1#2#3 {{\it Nucl. Phys. }{\bf #1} #2 {(#3)}}
\def\MPL#1#2#3 {{\it Mod. Phys. Lett. }{\bf #1} #2 {(#3)}}
\def\NC#1#2#3 {{\it Nuov. Cim. }{\bf #1} #2 {(#3)}}
\def\PREP#1#2#3 {{\it Phys. Report }{\bf #1} #2 {(#3)}}
\def\PROG#1#2#3 {{\it Prog. Theor. Phys. }{\bf #1} #2 {(#3)}}

\def\sq2{{1\over{\sqrt{2}}}}

\def\Dbar{\overline D}
\def\Bbar{\overline B}
\def\Kbar{\overline K}

\def\g5{\gamma_5}

\renewcommand{\Im}{\mbox{Im}\,}

\newcommand{\BR}{\,\mbox{BR}}

\def\beq{\begin{equation}}
\def\eeq{\end{equation}}
\def\be{\begin{equation}}
\def\ee{\end{equation}}
\def\ba{\begin{eqnarray}}
\def\ea{\end{eqnarray}}
\def\nn{\nonumber}
\def\bec{\begin{center}}
\def\eec{\end{center}}

\begin{document}

\Title{Search for New Physics in B Decays\footnote{Presented by A. I. Sanda}}
\bigskip

%
\label{AISAnda}

%
\author{ A. I. Sanda\index{Sanda} ~and  Tadashi Yoshikawa }

%
\address{Department of Physics\\
Nagoya University\\
Nagoya 464-8602, Japan\\
}

\makeauthor
\abstracts{
We review recent important progresses at the B factories and discuss the future prospects. We also comment on how we might proceed to search for new physics.
}

\section{Introduction}
Much progress in $B$ physics has been achieved over the past few years.
Both KEK and SLAC have achieved their corresponding design luminosity
goals, and are working hard to surpass them.
The $B\to\psi K_S$ asymmetry has been discovered. The direct CP asymmetry
has been discovered in the $B\to K\pi$ decay. According to the Belle result,
the $B\to \pi\pi$ CP asymmetry shows direct CP violation.
First measurements of $\phi_2$ and $\phi_3$ have been made as well as polarization
studies of $B\to \phi K^*,~\rho\rho,~\rho K^*$.

In this note, we shall review important $B$ factory results and then discuss
possibility for the upgrade.

\section{Selected achievements at $B$ factories}
\subsection{$\phi_1$}
\begin{figure}[b]
\centerline{
\includegraphics[height=5cm]{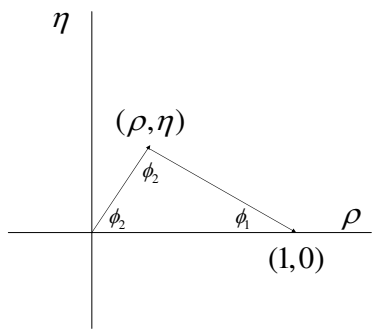}}

\caption{The unitarity triangle.
\label{fig:UT}}

\end{figure}

Who would have thought 5 years ago that we have a precision measurement
of CP asymmetry in $B\to\psi K_S$ decay?

The first angle of the unitarity triangle shown
in Fig. \ref{fig:UT} to be measured was $\phi_1$:
\ba
\sin 2\phi_1&=&+0.728\pm 0.056\pm 0.023 
~~~~~~~{\rm Belle}\cite{phi1},\nn \\
\sin 2\phi_1&=&+0.722\pm 0.040\pm 0.023 
~~~~~~~{\rm BABAR}\cite{phi1b}.
\ea
%
\begin{figure}[htb]
\centerline{
\includegraphics[height=7cm]{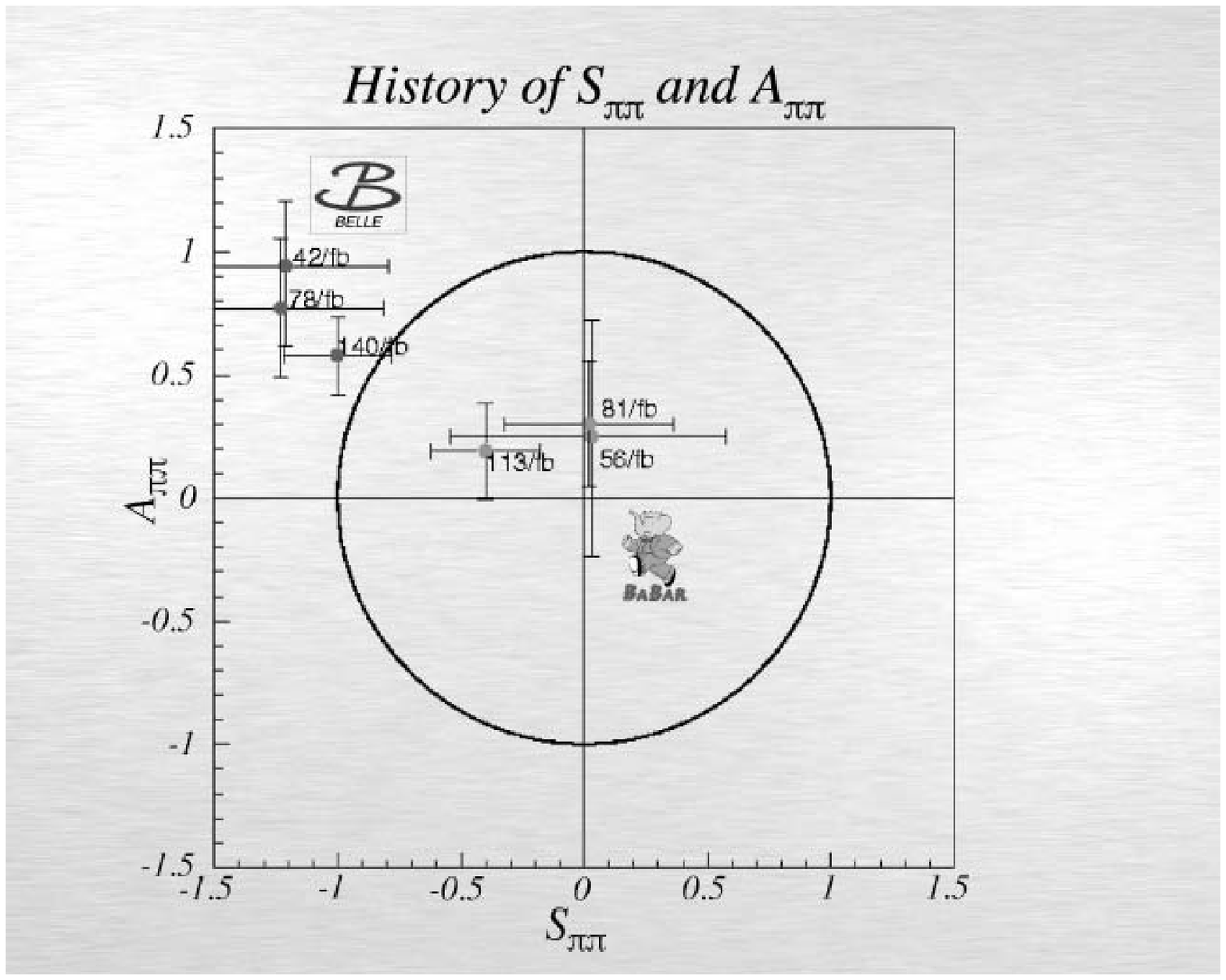}}
\caption{Belle and BABAR results on $B\to\pi\pi$ CP symmetry. The points on the upper left side represent Belle data
  and the points in the center represent BABAR data.
\label{fig:bd}}
\end{figure}

 The error is now less than 5\%. While this is certainly enough to
declare the correctness of the
Kobayashi-Maskawa theory, it is not enough, if we want to use this
information to look for New Physics beyond the Standard Model. 
It is worthwhile measuring it to the accuracy of 1\% as the 
theoretical uncertainty in relating this asymmetry to $\phi_1$ is of 
that order. 
\subsection{$\phi_2$}

The next challenge is $\phi_2$, but we are not so lucky here. We have both tree and
penguin amplitudes contributing to the  $B\to\pi\pi$ decay. Nevertheless, 
it is of great interest to pursue the time dependent CP asymmetry:

\beq
\frac{\Gamma_{\pi^+\pi^-}(t)-\overline\Gamma_{\pi^+\pi^-}(t)}
{\Gamma_{\pi^+\pi^-}(t)+\overline\Gamma_{\pi^+\pi^-}(t)}
=A_{\pi^+\pi^-}\cos(\Delta Mt)+S_{\pi^+\pi^-}\sin(\Delta Mt),
\eeq
where 
\beq
A_{\pi^+\pi^-}=\frac{|\bar\rho(\pi^+\pi^-)|^2-1}{|\bar\rho(\pi^+\pi^-)|^2+1}
~~~~~
S_{\pi^+\pi^-}=\Im\left(\frac{q}{p}\bar\rho(\pi^+\pi^-)\right).
\eeq
We can easily show that
\beq
|A_{\pi^+\pi^-}|^2+|S_{\pi^+\pi^-}|^2\le 1.
\eeq

Fig. 2 shows both Belle and BABAR results\cite{piilonen}. While it is tempting to say that the
direct CP violation in $B\to\pi\pi$ (non-vanishing $A_{\pi^+\pi^-}$) has been discovered at 
Belle, we feel that we should wait until their data comes within the circle.
Note that if it is established that the data point lies outside of the unit circle, it signals violation of quantum mechanics.

Both Belle and BABAR observe the $B\to\pi^0\pi^0$ decay:
\ba
Br(B\to\pi^0\pi^0)&=&(1.17\pm 0.32\pm 0.10)\times 10^{-6}
~~~~~~~{\rm BABAR}\cite{pi0b},\nn \\
Br(B\to\pi^0\pi^0)&=&(2.32\pm 0.48\pm 0.22)\times 10^{-6}
~~~~~~~{\rm Belle}\cite{pi0}.
\ea
This is very encouraging. Isospin analysis can be done. This may be a
place where $B$ factories continue to have the edge even after LHC turns on.
Certainly, Super $B$ luminosity should be defined to be that luminosity which gives 1\% measurement of $\phi_2$ using the isospin analysis.

\subsection{$\phi_3$}

The next challenge is $\phi_3$. One of the most promising ways is to
make use of the fact that we can not tell whether the intermediate state
is $D^0K^\pm$ or $\bar D^0K^\pm$ when we observe $D,\bar D\to K_S\pi\pi$ decay products in the final state:
\ba
B^\pm&\to& D^0K^\pm\to K_S\pi\pi K^\pm,\nn \\
B^\pm&\to& \bar D^0K^\pm\to K_S\pi\pi K^\pm.
\ea
Then amplitudes for these decays interfere, generating CP violation. This method was first suggested in Ref.\cite{bigisanda}. 

First results have been obtained:
\ba
\phi_3&=&(77^{+17}_{-19}(stat)\pm 13(syst)\pm
11(model))^\circ~~~~~~~{\rm Belle}\cite{DK},\nn \\
\phi_3&=&(88\pm 41(stat)\pm 19(syst)\pm 10(model))^\circ~~~~~~{\rm BABAR}\cite{DKb}.
\ea
Future progress in this method seems very promising.
We are getting into an era where we are starting to get results on the angles of the unitarity triangle. 
 We should compute the
required luminosity for the $B$ factory upgrade based on a 1\% determination of 
$\phi_2$ and $\phi_3$.

\subsection{Direct CP asymmetries in $K\pi$ }
Large direct CP asymmetry in $B\to K\pi$ decay 
has been predicted 
in the PQCD method and it has been observed in:
\beq
\frac{\BR(\Bbar\to K^-\pi^+)-\BR(B\to K^+\pi^-)}
{\BR(\Bbar\to K^-\pi^+)+\BR(B\to K^+\pi^-)}
=-0.113\pm 0.019
\eeq
An asymmetry of similar size has been predicted 
in $B^\pm\to K^\pm\pi^0$ but actual measurement
shows that:
\beq
\frac{\BR(B^-\to K^-\pi^0)-\BR(B^+\to K^+\pi^0)}
{\BR(B^-\to K^-\pi^0)+\BR(B^+\to K^+\pi^0)}
=0.04\pm 0.04\nn
\eeq
Theoretically, the fact that these asymmetries must be equal follows 
rather generally if the color suppressed amplitudes
and electroweak penguin diagrams are small.
Experimental measurement shows that these amplitudes are not negligible, and that they play an important role. If these amplitudes are important they may
also modify $B\to\pi^0\pi^0$ decay rate. Details of
this type of analysis has been presented by Yoshikawa.

\section{New Physics searches}
\subsection{$B\to\phi K_S$}
In the Standard Model (SM), the amplitudes for $B\to\psi K_S$ and $B\to\phi K_S$
have equal phases. So, we expect $S_{\phi K_S}=S_{\psi K_S}=\sin(2\phi_1)$.
But, Belle obtained\cite{sakai}:
\ba
S(\phi K^0)&=&+0.06\pm 0.33 \pm 0.09,\nn \\
A(\phi K^0)&=&+0.08\pm 0.22 \pm 0.09.
\label{S}
\ea
Note that the Belle result for $S(\phi K^0)$ is dramatically different
from the previous result, $S(\phi K^0)=-0.96\pm 0.5^{+0.09}_{-0.06}$\cite{HFAG}. This is due to the fact that
their new measurement with new vertex detector yielded  $S(\phi K^0)=+0.78\pm 0.45$. Averaging all the data, they obtained the value shown in (\ref{S}). While the data taken with the new vertex detector 
yields roughly the result $\sim \sin(2\phi_1)$, as
expected from the SM, and a Monte Carlo study shows
that the probability for this sign flip-flop is about
4.5\%, it is nevertheless mind
boggling.

The result of Eq. (\ref{S}) is off from the 
SM prediction by about $2.2 \sigma$. One of the authors AIS) is reminded of what
Professor Wong-Young Lee told him once when he was a post doc at Columbia.
He said, ``A $3\sigma$ effect goes away half of the time!" So, we would wait
until there is more convincing data before we tell ourselves that New Physics
has been discovered. 
But, depending on the confidence of the experimentalists,
this discrepancy should be a major motivation for building the
$B$ factory upgrade.

\begin{figure}[t]
\vspace{-3cm}
\centerline{
\includegraphics[height=10cm]{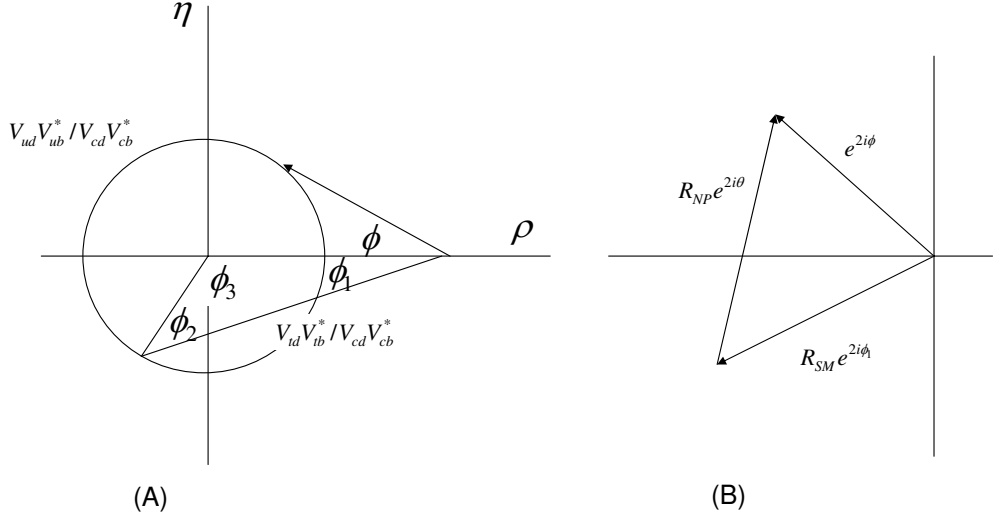}}
\caption{Suppose there is NP which contributes to $M_{12}$. The experimental measurement of the $\psi K_S$ asymmetry gives $\phi$ as shown in (A). Let us 
entertain an extreme situation where $\phi_3$ is negative. Then the unitarity triangle is located 
below the horizontal axis as shown here. Figure (B) shows the relationship (see Eq. (9)) between the vector representing the observed asymmetry, $e^{2i\phi}$, 
the vector representing NP, $R_{NP}e^{2i\theta}$, and the vector representing the SM
 contribution, $R_{SM}e^{2i\phi_1}$. 
\label{fig:RNP}}

\end{figure}

\subsection{Dilepton asymmetries in $B\Bbar\to l^\pm l^\pm +anything$}
If there is New Physics (NP), it might first show up in $\Delta M$.
Obviously, when we search for NP, the SM contribution is
the background. Since $\Delta M$ is of the second order in the weak 
interaction, it may 
be easier to observe NP contributions here.
We define\cite{xing}:
\beq
M_{12}=M_{12}^{SM}+M_{12}^{NP}\equiv\frac{\Delta M}{2}
\left(R_{SM}e^{2i\phi_1}+R_{NP}e^{2i\theta}\right)
=\frac{\Delta M}{2}e^{2i\phi},
\eeq
where $M_{12}^{NP}$ is the NP contribution to $M_{12}$.
The dilepton CP asymmetry is given by\cite{bigisanda2}:
\ba
A_{SL} \equiv \frac{N^{++}-N^{--}}{N^{++}+N^{--}}
     &=&\Im\frac{\Gamma_{12}}{M_{12}}\nn \\
&=&r~\Im\left(\frac{V_{ub}V^*_{ud}+V_{cb}V^*_{cd}}{V_{tb}V_{td}^*+\frac{R_{NP}}{R_{SM}}|V_{tb}V_{td}^*|e^{2i\theta}}\right)^2+{\cal O}\left(r\frac{m_c^2}{m_b^2}\right),
\ea
where $r={\cal O}(10^{-3})$ is computed in the SM.
If $R_{NP}$ is not present, the unitarity constraint
of the KM matrix forces the leading term to vanish and the asymmetry is
${\cal O}\left(r\frac{m_c^2}{m_b^2}\right)$. 
%
%

The actual computation of $\frac{\Gamma_{12}}{M_{12}}$ may be tricky as it may receive substantial contribution from long distance effects. Here we assume that contributions from intermediate states with $\alpha \beta$ ($\alpha,\beta=u~{\rm or}~c$) quarks appropriately average the long distance effects, and give a sufficiently good approximation.
The fraction $\frac{\Gamma_{12}}{M_{12}}$ has been computed including the next to leading order QCD corrections\cite{BBLN}.
Write contribution to $\frac{\Gamma_{12}}{M_{12}}$ from the box diagram
where the inner lines are $(\alpha,\beta)$ quarks as 
\be
F_{12}^{\alpha\beta}(V_{\alpha b}V_{\alpha d}^*)(V_{\beta b}V_{\beta d}^*).
\ee
Then the result is given as\footnote{The actual expression for $F_{12}^{\alpha\beta}$ is given in Ref.\cite{BBLN}}:
\ba
\frac{\Gamma_{12}}{M_{12}} 
  &=& \frac{(V_{tb}V_{td}^*)^2}{M_{12}^{SM}}
         \left[-F^{cc}_{12}+2(F^{uc}_{12}-F^{cc}_{12})
                 \frac{V_{ub}V_{ud}^*}{V_{tb}V_{td}^*}
            +(2F^{uc}_{12}-F^{cc}_{12}-F^{uu}_{12})
                  \frac{(V_{ub}V_{ud}^*)^2}{(V_{tb}V_{td}^*)^2}\right],
\label{G12/M12}
\ea
The dilepton CP asymmetry is written as a function of $\phi_1$ as follows:
\ba 
A_{SL}  
       &=& Im \{ \frac{\Gamma_{12}}{M_{12}^{SM}} \} R_{SM} \cos2(\phi - \phi_1)
         - Re \{ \frac{\Gamma_{12}}{M_{12}^{SM}} \} R_{SM} \sin2(\phi
       - \phi_1)
\label{asl}
\ea
The KM factors in $\Gamma_{12}/M_{12}^{SM}$ and $R_{SM}$ can be also 
written as the functions of $\phi_1$. In the SM, $\phi_1$ should be
same with $\phi $ which is measured by the CP asymmetry in
$B\rightarrow \psi K_s $ so that the contribution is only the
first term in Eq. (\ref{asl}) and comes from the imaginary part of 
the second and third terms in Eq. (\ref{G12/M12}), which vanishes in the limit $m_u=m_c$.
The SM contribution is roughly $10^{-4}$.  
The presence of $R_{NP}$ spoils the cancellation and
the second term in Eq. (\ref{asl}) becomes non-vanishing. In this case, the CP asymmetry may become as large as a few \%.

If this asymmetry is measured to be much larger than
 ${\cal O}(10^{-4})$, it implies the presence of NP. 
The best limit on this asymmetry is given by Belle\cite{dilepton}:
\beq
\frac{N^{++}-N^{--}}{N^{++}+N^{--}}=(-0.13\pm 0.60 \pm 0.56)\%.
\eeq
It is interesting to note that
$M_{12}^{NP}$ does not have to be complex. The presence of $R_{NP}$,
which means there may be a difference between $\phi_1 $ and $\phi $, 
spoils the cancellation of the KM phase, leading to the asymmetry.

\begin{figure}[h]
\centerline{
\rotatebox{0}{\includegraphics[height=5cm]{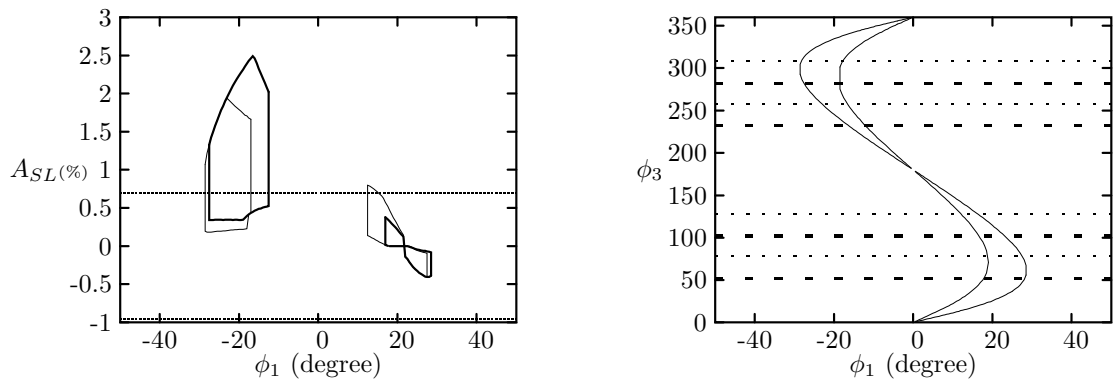}}}
\caption{ In the left figure 
the allowed region of the dilepton CP asymmetry $A_{SL}$
for $\phi_1$ as the angle of $V_{tb}$ in the SM  
by taking account of the constraint $\phi_3 = 77^\circ \pm 25^\circ 
$\cite{DK} and in the right figure the allowed region for $\phi_3$ are
plotted. The dotted lines show the bounds from experimental data of
$\phi_3$ with four fold ambiguity. The regions by thick (thin) line in the
left correspond to the bounds shown by thick(thin) dotted lines in the
right one. The dotted line in left figure
shows the experimental bound of $A_{SL}$ by Belle.  }\label{fig:ASL}
\end{figure}

In Fig. 3(A) we show an example of how the $\rho-\eta$ plot gets modified by a non-vanishing $R_{NP}$. The
CP asymmetry in $B\to\psi K_S$ determines $\phi$.
For an illustration, let us consider a remote
possibility that $\phi_3$ turned out to be negative.
Then we have a situation depicted in this figure. Fig. 3(B) gives the required 
$R_{NP}e^{i\theta}$.

In Fig. \ref{fig:ASL}, the allowed
region for $A_{SL}$ can be shown in terms of $\phi_1$ under the 
constraint of $\phi_3$. But there are four fold ambiguity to measure
$\phi_3$ and the experimental bounds from $\phi_3= 77^\circ \pm 25^\circ
$\cite{DK} with the ambiguity are
plotted by dotted lines in the right figure.  
Under taking account of the constraint of $\phi_3$ for $\phi_1$,  $A_{SL}$ is
plotted in the left figure.  
The region by thick(thin) line in left figure is from the constraints
for $\phi_3$ by thick(thin) dashed line in right figure. 
These figures may tell us that
combining the constraints from $A_{SL}$ and $\phi_3$ can reduce 
the parameter space for NP and more accurate measurement will help to
solve the ambiguity for $\phi_3$.      
 Further improvement of the
upper limit is strongly encouraged.  
	 

\subsection{Lepton number violation}
We now know that there is neutrino mixing - lepton flavor number is violated.
This may show up in $\tau\to e\gamma$, $\tau\to \mu\gamma$, $\tau\to 3\mu$
$\tau\to 3e$, $\tau\to e\mu\mu$, etc. Belle has already obtained the following 90\% CL limits\cite{leptonv}:
\ba
Br(\tau\to \mu\gamma)&<&3.1\times 10^{-7},\nn \\
Br(\tau\to e\gamma)&<&3.8\times 10^{-7}.
\ea
It is not so unrealistic to expect that these lepton number violating 
processes are actually observed in the near future.

It has been customary to study quark physics and lepton physics separately. Since we found that the lepton number is not conserved, it is perhaps advantageous
to study the quark system and the lepton system
in an unified manner. Searching for lepton number violation 
in $B$ decays, 
such as $B\to\tau\mu$ and $B\to 3\mu$, is good example of this unification.

\begin{table}[h]
\caption{Examples of lepton number violating decays. Lepton number violation may very well show up in $B$ decays.} \vspace{1mm}
\small
\begin{center}
\begin{tabular}{|c|c|}
\hline 
{\rm Quark~ physics}&$B\to\tau\mu,~ B\to 3\mu,$~ etc.\\
\hline
{\rm Lepton ~physics}&$\mu\to e\gamma,~ \tau\to 3\mu,$~ etc. \\
\hline 
\end{tabular}
\vspace{-1mm}
\end{center}
\end{table}

\section*{Conclusion}

Much exciting flavor physics with B and $\tau$ decays remains uncovered. We hope that Belle and Babar come to an agreement on $A_{\pi\pi}$ and $S_{\pi\pi}$ measurements. This should be followed by first results on the 
isospin analysis for $B\to\pi\pi$ decays. Theoretical understanding of CP asymmetry for $B^\pm\to K^\pm\pi^0$ decay must be achieved. It is likely that the CP
asymmetry for $B\to\phi K_S$ will show new physics
at the level of less than 5\% as opposed to 50-100\%
level. Dilepton asymmetry will put nontrivial constraints on new physics in the near future. Lepton number violation may be around the corner.

\section*{Acknowledgments}

 We acknowledge support from the Japan Society for the Promotion of Science,
Japan-US collaboration program, and a grant from Ministry of Education, Culture, Sports,
Science and Technology of Japan. The work of T.Y. was supported by 21st Century COE Program of Nagoya University.


\end{document}